# On the Deutron Relativistic Wave Function[*]

## L. Abesalashvili[1,**], Akhobadze[1], V. Garsevanishvili[2], T. Jalagania,[1] Yu. Tevzadze[1]

[1] Institute of High Energy Physics, Tbilisi State University, 9, University Str. 0186 Tbilisi, Georgia.
[2] A. Razmadze Mathematical Institute, 1, Alexidze Str. 0193 Tbilisi, Georgia.

Date: February 15, 2011

**Abstract**—Light front form of the relativization of the deuteron wave function is considered. Parametars of the wave function are extracted comparing theoretical results with experimental data. Experimental data are obtained on the two-metre propane bubble chamber of JINR (Dubna) bombarded by the deutron beam with momentum of 4.2 GeV/c/nucleon.

PACS numbers: 13.75.Cs, 13.85.-t, 21.30.-x, 24.85.+t, 25.70.-z

From the very beginnig of the development of the relativistic nuclear physics a problem of relativization of nuclear wave functions has been posed. The problem is twofold: relativization of the intrinsic motion of nucleons in nuclei and the relativization of the movement of the nucleus as a whole [1].

In the present note relativization of the deuteron wave function is considered. From the experimental point of view the most convenient is the study of stripping proton distribution in the bubble chamber (the incoming nucleus is deuteron), since the deuteron relativistic wave function enters this distribution as a multiplier.

Consider the relativistic deuteron break up on the carbon target, when stripping protons are detected in the final state. In the impulse approximation the cross section of this process looks as follows [2]:

$$E^{st}\frac{d\sigma}{d\vec{p}} \sim \frac{\lambda^{1/2}(s_{NN},m^2,m^2)}{\lambda^{1/2}(s,m^2,m_d^2)} \sigma_{tot}(s_{NN}) \left| \frac{\Phi(x,\vec{p_\perp})}{1-x} \right|^2 \qquad (1)$$

Here s is the usual Mandelstam variable,

$$s_{NN} = s(1-x^{st}) + m^2 - \frac{\vec{p_\perp^{st^2}}}{x^{st}} + \frac{m^2}{x^{st}} \qquad (2)$$

$\sigma_{tot}(s_{NN})$ is nucleon-nucleon total cross section, m is the nucleon mass, $m_d$ is the deuteron mass.

$$\lambda(x,y,z) = (x-y-z)^2 - 4yz \qquad (3)$$

---

[*] The text was submitted by the authors in English.

[**] E-mail: abesal@hepi.edu.ge; lianaabesalashvili@yahoo.com

The variable $x^{st}$ in the laboratory frame is defined as follows:

$$x^{st} = \frac{E^{st} + p^{st}}{(E_d + p_d) + m} \quad (4)$$

$\Phi(x, \vec{p_\perp})$ is the deuteron relativistic wave function. Its arguments are related to the observable quantities as follows:

$$x = 1 - (1 + \frac{m}{E_d + p_d})x^{st} \quad (5)$$

$$\vec{p_\perp} = -\vec{p_\perp}^{st} \quad (6)$$

We choose the deuteron relativistic wave function as a generalization of the well – known Hulthen wave function:

$$\Phi(x, \vec{p_\perp}) \sim \frac{1}{[\frac{\vec{p_\perp}^2 + m^2}{x(1-x)} - \alpha_R][\frac{\vec{p_\perp}^2 + m^2}{x(1-x)} - \beta_R]} \quad (7)$$

$\alpha_R$ and $\beta_R$ are adjustable relativistic parameters, which are extracted from the experimental data.

From (1) one dimensional distributions $\frac{d\sigma}{dx^{st}}$ and $\frac{d\sigma}{dp_\perp^{st}}$ are obtained:

$$\frac{d\sigma}{dx^{st}} = \int_0^{P_{\perp max}^{st}} \frac{d\sigma}{dx^{st} dp_\perp^{st}} dp_\perp^{st} \quad (8)$$

$$\frac{d\sigma}{dp_\perp^{st}} = \int_{x_{min}^{st}}^{x_{max}^{st}} \frac{d\sigma}{dx^{st} dp_\perp^{st}} dx^{st} \quad (9)$$

Characteristic feature of the $x^{st}$ distribution is the existence of maximum at the point:

$$x^{st} = \frac{1}{2(1 + \frac{m}{E_d + p_d})} \quad (10)$$

Experimental data are obtained on the 2 metre propane bubble chamber *PBC-500* in the Laboratory of High Energies of the Joint Institute of Nuclear Research ( Dubna ).The chamber was bombarded by deuteron beam with momentum of 4.2 GeV/c per nucleon. Methodic problems of experiment are given in Refs [3-8]. Proton is called to be stripping if its momentum is bigger than 3 GeV/c an emission angle less than 4 º .

**L. Abesalashvili, L. Akhobadze, V. Garsevanishvili, T. Jalagania,Yu.Tevzadze**

In Fig. 1 $\frac{d\sigma}{dx^{st}}$ distribution of stripping protons is compared with corresponding experimental data and parameters $\alpha_R$ and $\beta_R$ are extracted.

In Fig. 2 the same procedure is performed for $\frac{d\sigma}{dp_\perp^{st}}$ distribution.

Numerical values of the parameters $\alpha_R$ and $\beta_R$ are in a good agreement with the results, obtained by other methods [2].

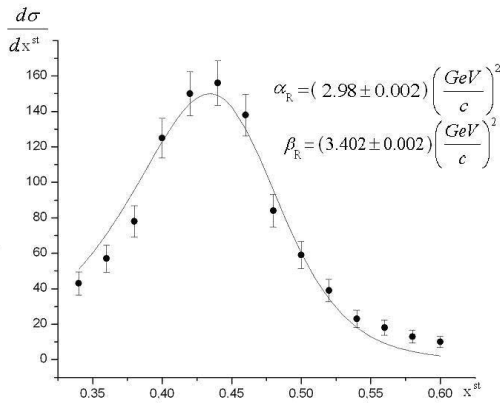 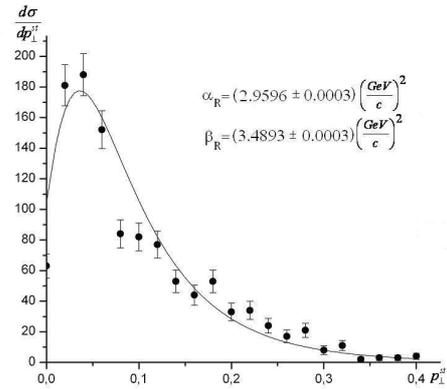

Fig. 1. $\frac{d\sigma}{dx^{st}}$ distribution of stripping protons from the reaction $dC \to p^{st} + X$

Fig. 2. $\frac{d\sigma}{dp_\perp^{st}}$ distribution of stripping protons from the reaction $dC \to p^{st} + X$

This work is supported by the Georgian National Science Foundation (grant GNSF/ST08-418).

**L. Abesalashvili, L. Akhobadze, V. Garsevanishvili, T. Jalagania, Yu. Tevzadze**

**О релятивистской волновой функции дейтрона**

Рассмотрена релятивизация волновой функции дейтрона в формализме светового фронта. Параметры волновой функции установлены из сравнения теоретических результатов с экспериментальными данными. Экспериментальные данные получены на двухметровой пропановой камере ОИЯИ (Дубна), бомбардируемой пучком дейтронов с импульсом 4.2 GeV/c на нуклон.

**L. Abesalashvili, L. Akhobadze, V. Garsevanishvili, T. Jalagania,Yu.Tevzadze**